\documentclass[12pt]{article}
\usepackage{epsfig,cite,graphicx}
\begin{document}
\begin{center}
%%%%%%%%%%%%%%%%%%%%%%%%%%%%%%%%%%%%%%%%%%%%%%%%%%%%%%%%%%%%%%%%%%%%%%%%%%%
{\bf\Large Hypercentral Constituent Quark Model with a Meson Cloud}\\
%%%%%%%%%%%%%%%%%%%%%%%%%%%%%%%%%%%%%%%%%%%%%%%%%%%%%%%%%%%%%%%%%%%%%%%%%%
\vspace*{1cm}
%%%%%%%%%%%%%%%%%%%%%%%%%%%%%%%%%%%%%%%%%%%%%%%%%%%%%%%%%%%%%%%%%%%%%%%%%%%
D. Y. Chen $^{1}$, Y. B. Dong$^{2,1}$, M. M. Giannini$^{3}$ and E.
Santopinto$^{3}$ \\
\vspace{0.3cm} %
{Institute of High Energy Physics \\
Chinese Academy of Science,\ Beijing,\ 100049,\ P.R.China$^1$}\\
{CCAST (World Lab.) Beijing 100080, P. R. China$^2$}\\
and\\
{INFN, Sezione di Genova and Dipartimento di Fisica dell
Universit\`a di Genova \\ Genova,
Italy$^3$ }
%%%%%%%%%%%%%%%%%%%%%%%%%%%%%%%%%%%%%%%%%%%%%%%%%%%%%%%%%%%%%%%%%%%%%%%%%%%
\vspace*{1cm}
\end{center}
%%%%%%%%%%%%%%%%%%%%%%%%%%%%%%%%%%%%%%%%%%%%%%%%%%%%%%%%%%%%%%%%%%%%%%%%%%%
\begin{abstract}
The results for the elastic nucleon form factors and the
electromagnetic transition amplitudes to the $\Delta(1232)$
resonance, obtained with the Hypercentral Constituent Quark Model
with the inclusion of a meson cloud correction are briefly
presented. The pion cloud effects  are explicitly discussed.\\
\end{abstract}

%%%%%%%%%%%%%%%%%%%%%%%%%%%%%%%%%%%%%%%%%%%%%%%%%%%%%%%%%%%%%%%%%%%%%%%%%%%

\section{Introduction}
\label{introduction}%
In recent years, various constituent quark potential models have
been proposed to describe the intrinsic structure of baryons.
Typical examples of the models are the Igsur-Karl model
\cite{isgur}, the Capstic and Isgur relativized model
\cite{capstic}, the chiral model \cite{glozman} and the Hypercentral
Constituent Quark Model (HCQM) \cite{PLB364}. Various baryon
properties, such as baryon spectroscopy, nucleon form factors and
the transverse and the longitudinal electromagnetic transition
amplitudes of $3$ and $4$ stars resonances, have been systematically
calculated with the HCQM \cite{PLB364, JPG24, PRC62}. In this work,
we will present calculations of the nucleon electromagnetic form
factors and nucleon-$\Delta(1232)$ electromagnetic transition
amplitudes based on the framework of the Hypercentral Constituent
Quark Model with a meson cloud. Particularly, we shall stress the
corrections due to the pion meson cloud. In our calculation, a
baryon is considered as a three-quark core surrounded by the pion
meson cloud. Thus, we have new degrees of freedom in addition to the
conventional three constituent quarks.
%%%%%%%%%%%%%%%%%%%%%%%%%%%%%%%%%%%%%%%%%%%%%%%%%%%%%%%%%%%%%%%%%%%%%%%%%%%

\section{The Hypercentral Model}
\label{Hyc model}%
The Hypercentral Constituent Quark Model \cite{PLB364} contains  a
linear plus a coulomb-like potential
\begin{eqnarray}
V(x)=-\frac{\tau}{x}+\alpha x, ~~~~ \mbox{with}~~~~~~
x=\sqrt{\mbox{\boldmath{$\rho$}}^2+\mbox{\boldmath{$\lambda$}}^2}
\label{vx}
\end{eqnarray}
where $x$ is the hyperradius defined in terms of the standard Jacobi
coordinates \mbox{\boldmath{$\rho$}} and
\mbox{\boldmath{$\lambda$}}. It should be mentioned that this
hypercentral potential has the following features. First of all, it
contains three-body force effects. Secondly, it can be considered as
the hypercentral approximation of a two-body potential of the form
linear plus coulomb, as suggested by lattice QCD calculations
\cite{latgunnar}. Thirdly, its predictions of the proton form
factors decrease as powers of the virtual photon momentum, while the
form factors of the conventional harmonic oscillator potential
decrease as gaussians. Finally, its predictions for the transition
amplitudes $A_{1/2}(Q^2)$ of $S_{11}(1535)$ and $D_{13}(1520)$
resonances \cite{JPG24} agree with the data, particularly in the
momentum transfer region of $Q^2\sim 1\sim 1.5GeV^2$.
\par  %
The hypercentral  model contains also a standard hyperfine interaction
fitted to the $N-\Delta$ mass difference and in this form it
has been employed  for the calculation of various baryon properties
\cite{JPG24,PRC62, PLB387}. However it has been shown that a good description
of the lower part of the spectrum and of the transition amplitudes
can be obtained also using simply the hyper-Coulomb (Hyc) potential to
which a linear and a spin-dependent interactions are added as perturbations
\cite{ref2,r121}.

Thus, for the low-lying resonance states it is a good approximation to
simply take  the space wave functions of the Hyc potential instead of the
ones coming from the numerical solution of  the linear plus Coulomb
potentials with hyperfine interaction.
\par%

\section{Pion meson cloud correction}
\label{correction}%
To take the pion meson cloud effects into account, one may consider
the following Lagrangian density with a $\pi qq$ coupling ( similar to
the $\pi NN$ case) \cite{NPA393}
\begin{eqnarray}
\mathcal{L}&=&
i\overline{\psi}_q(x)\gamma^{\mu}\partial_{\mu}\psi_q(x)-
m_q\overline{\psi}_q(x)\psi_q(x)+
\frac{1}{2}(\partial_{\mu}\pi(x))^2-\frac{1}{2}m_{\pi}^2\pi^2(x)\nonumber\\
&&-ig\overline{\psi}_q(x)\gamma_5\psi_q(x)\mathbf{\tau}\cdot\mathbf{\pi}(x).
\label{lag}
\end{eqnarray}
From Eq. (\ref{lag}) the conserved local electromagnetic current can
be derived using the principle of minimal coupling
$\partial_\mu\rightarrow\partial_\mu+i e_q A_\mu$, where $e_q$ is the
charge carried by the field upon which the derivative operator acts.
The total electromagnetic current $J^\mu$ is then
\begin{eqnarray}
J^\mu(x)&=&j_q^\mu(x)+j_\pi^\mu(x),
\end{eqnarray}
where
\begin{eqnarray}
j_q^\mu(x)&=& \sum_a Q_a e\overline{\psi}_a(x)\gamma^\mu
\psi_a(x),\nonumber\\
j_\pi^\mu(x)&=&-ie[\pi^\dag(x)\partial^\mu\pi(x)
-\pi\partial^\mu\pi(x)^\dag(x)].
\end{eqnarray}
Because of the $\pi qq$ coupling, a physical baryon state can be
described as a superposition of a three-quark core and its surrounding pion
cloud,

\begin{eqnarray}
\left|A\right>=\sqrt{Z_2^A}\left[1+(E_A-H_0-\Lambda
{\mathcal{H}}_{int} \Lambda)^{-1}{\mathcal{H}}_{int}\right]
\left|A_0\right> ,
\label{z2a}    %
\end{eqnarray}
where $Z_2^A$ is the bare baryon probability in the physical baryon
state, $\Lambda$ is the projection operator projecting  out all the
components of $\left|A\right>$ with at least one pion, and
${\mathcal{H}}_{int}$ is the interaction Hamiltonian which describes
the process of emission and absorption of pions, which can be
obtained from the Lagrangian density with the $\pi qq$ coupling,
Eq. (\ref{lag}) \cite{NPA393}.

\par%

Our numerical calculations of the elastic nucleon form factors and
of the $N-\Delta$ electromagnetic transition amplitudes are
performed with the analytical  model of Ref. \cite{ref2,r121}
without the spin-spin interaction and with the explicit  inclusion
of the pion cloud corrections.  We set the parameters for the HCQM
potential as $\tau=4.59$, $\alpha=1.61 fm^{-2}$ \cite{PLB364} and
$g=0.585$ (corresponding to the usual $\pi NN$ coupling constant
$g^2_{\pi NN}/(4\pi)=13.6$).

\par%

Moreover, we also take the results of the conventional harmonic
oscillator potential with $\alpha=0.410~GeV$ (corresponding to a
wave function with radius of the order of $0.5~fm$) for a
comparison. To calculate the electromagnetic interaction between the
photon and the nucleon with the pion meson cloud, we consider the following
couplings: photon-quark, photon-quark with the pion meson cloud in
flight, and  photon-charged pion. Fig. 1 illustrates the three
couplings between the photon and the nucleon or the $\Delta$ with the pion
meson cloud.

\par%

%%%%%%%%%%%%%%%%%%%%%%%%%%%%%%%%%%%%%%%%%%%%%%%%%%%%%%%%%%%%%%%%%%%%%%%%%

\begin{figure}[h]

\centering

\includegraphics[width=150mm]{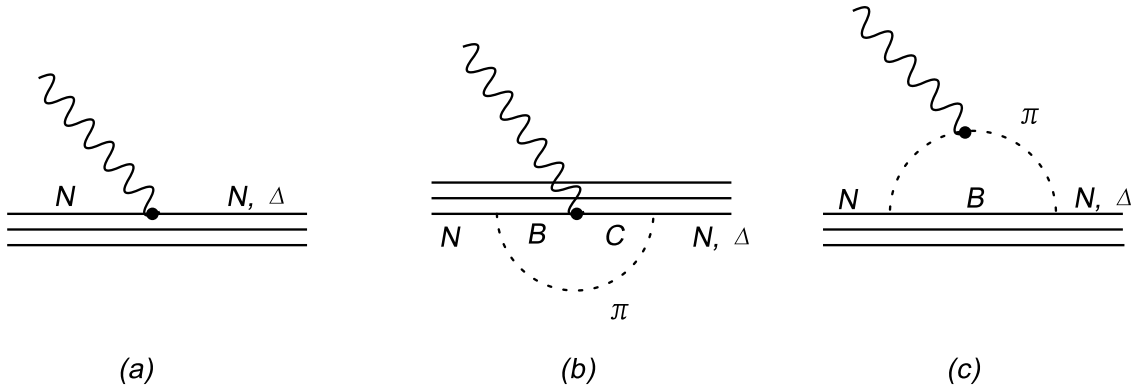}  %
\renewcommand{\figurename}{Fig.}
\caption{Diagrams illustrating the various contributions included in
the calculation. The intermediate baryons $B$ and $C$ are here restricted
to the $N$ and $\Delta$.} %
\label{feynman}
\end{figure}

%%%%%%%%%%%%%%%%%%%%%%%%%%%%%%%%%%%%%%%%%%%%%%%%%%%%%%%%%%%%%%%%%%%%%%%%%

%%%%%%%%%%%%%%%%%%%%%%%%%%%%%%%%%%%%%%%%%%%%%%%%%%%%%%%%%%%%%%%%%%%%%%%%%%

\begin{figure}[h]
\centering
\includegraphics[width=75mm,height=75mm]{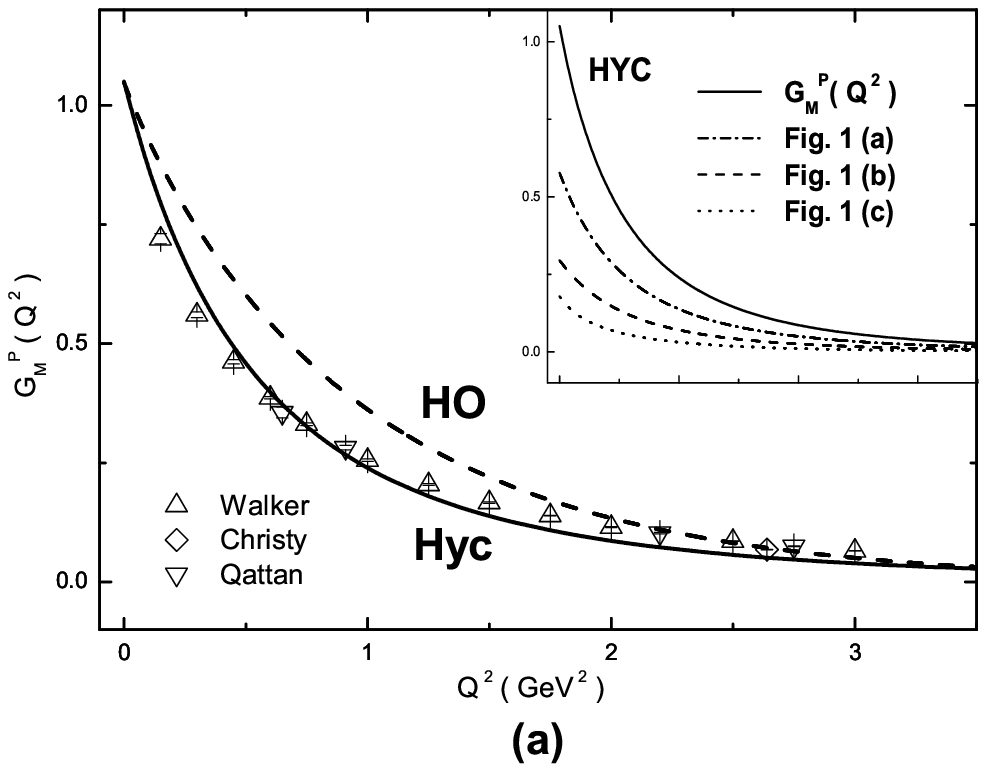}  %
\hspace{7mm} %
\includegraphics[width=75mm,height=75mm]{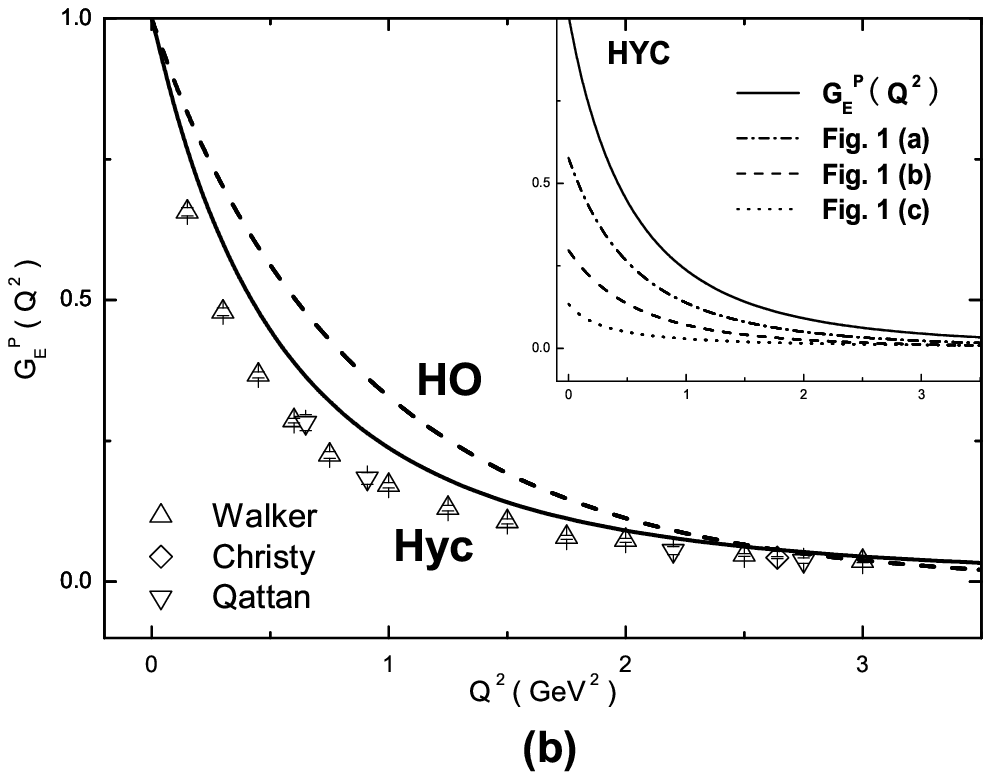}  %
\renewcommand{\figurename}{Fig.}
\caption{Comparison between the experimental data and  the
theoretical predictions (Hyc and HO) for (a) the proton magnetic
form factor and (b) the proton electric form factor. The
experimental data are taken from \cite{pexp}. Each figure also
explicitly shows the different contributions from Fig. \ref{feynman}
for the Hyc case.}
\label{pf} %
\end{figure}

%%%%%%%%%%%%%%%%%%%%%%%%%%%%%%%%%%%%%%%%%%%%%%%%%%%%%%%%%%%%%%%%%%%%%%%%%%
\section{Results and discussions}
\par\noindent\par%

As shown in Eq. \ref{z2a}, a baryon wave function contains
a three-quark core component and a three-quark core plus a pion
meson cloud component. One can directly calculate the three-quark
core probability $Z_2^A$\cite{NPA393}. For the nucleon, we have
$Z_2^N=0.567$ in the Hyc potential and $Z_2^N=0.442$ in the classical HO
potential. Our result for  $<r>_P$ is $0.68~ fm$ in Hyc potential and
$0.54~ fm$ in the HO potential. The probability of the pion meson in the HO
potential is larger than that in the Hyc potential. This can be explained
by the fact that the pionic contribution is competing with that of the  quark
core and a smaller r.m.s radius means a stronger pion coupling.

\par%

Figure \ref{pf} shows the obtained proton electromagnetic form
factors compared with the experimental data.  We use the Hyc wave
function as the wave function of the nucleon. Figure \ref{pf} also
shows different contributions from Fig. 1 to the proton form factors
(see the insets in the top right corner of Figs.(2a) and (2b)). In
the small $Q^2$ region, the contribution of the $\gamma \pi\pi$
interaction is more than $15\%$ and  as $Q^2$ increases it decreases
quickly. When $Q^2$ is beyond $1.0~GeV^2$ the contribution nearly
vanishes. One finds that the contribution of the pion meson cloud is
important especially in the low $Q^2$ region and also one can draw
the conclusion that for the proton form factors the hypercentral
constituent quark model can give a better prediction than that of
the HO model.

\par%

%%%%%%%%%%%%%%%%%%%%%%%%%%%%%%%%%%%%%%%%%%%%%%%%%%%%%%%%%%%%%%%%%%%%%%%%%%
\begin{figure}[h]
\centering
\includegraphics[width=75mm,,height=75mm]{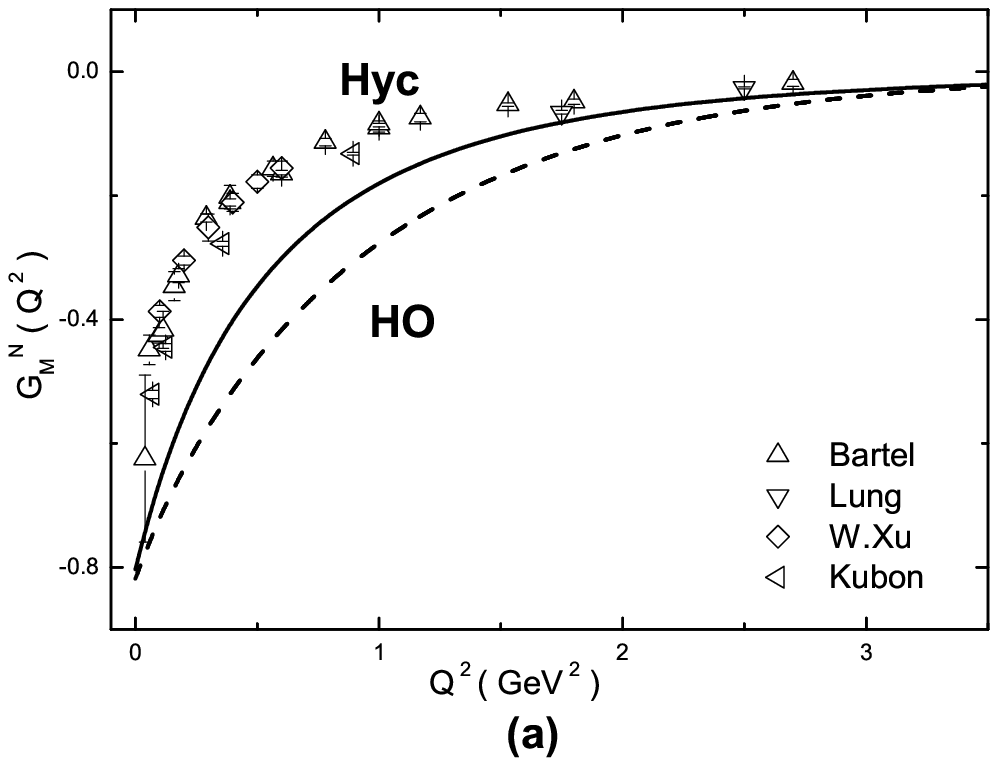}  %
\hspace{7mm}%
\includegraphics[width=75mm,,height=75mm]{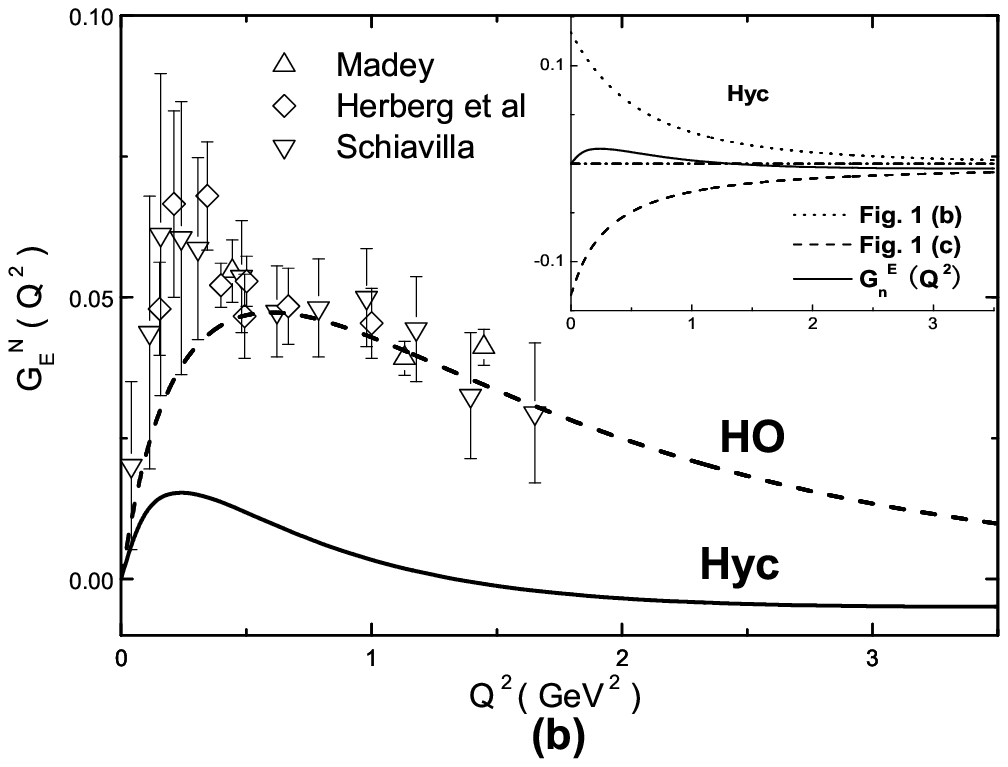}  %
\renewcommand{\figurename}{Fig.}
\caption{Neutron magnetic (a) and electric (b) form factors
predicted by the Hyc potential and the HO potentials compared with
the experimental data \cite{nmexp,neexp}. The inset in the top right
corner of (b) shows the different contributions to $G_E^N$ of Fig.
\ref{feynman} in the Hyc case}.
\label{nf}    %
\end{figure}

%%%%%%%%%%%%%%%%%%%%%%%%%%%%%%%%%%%%%%%%%%%%%%%%%%%%%%%%%%%%%%%%%%%%%%%%%%
Our results for the neutron electromagnetic form factors are shown in
Figure \ref{nf}. For the magnetic form factor the Hyc gives a
better prediction, however for the electric form factor the result
of the Hyc wave function is smaller than the experimental data.
It is also smaller than that of HO potential since the pionic contribution
in the Hyc potential is smaller and the non-zero value of the neutron
charge distribution in the present calculation results from the pionic
contribution.  It should be stressed that here we have not considered the
hyperfine mixing. We know that spin-spin forces lead to configuration mixing;
for example, the octet wave function $\psi(8,~\frac{1}{2}^+)$ not
only gets contribution from the octet $\left|56,~0^+\right>_{N=0}$
but also from $\left|56,~0^+\right>_{N=2},~\left|70,~0^+\right>_{N=0}$ and
$\left|70,~2^+\right>_{N=2}$. This mixing gives rise to a non-zero electric
form factor of the neutron.  It is expected that the mixing effect
has to be taken into account for a further analysis of the
neutron charge form factor in order to get a better result in comparison with
the experimental data

\par%

%%%%%%%%%%%%%%%%%%%%%%%%%%%%%%%%%%%%%%%%%%%%%%%%%%%%%%%%%%%%%%%%%%%%%%%%%%
\begin{figure}[h]
\centering
\includegraphics[width=75mm,height=75mm]{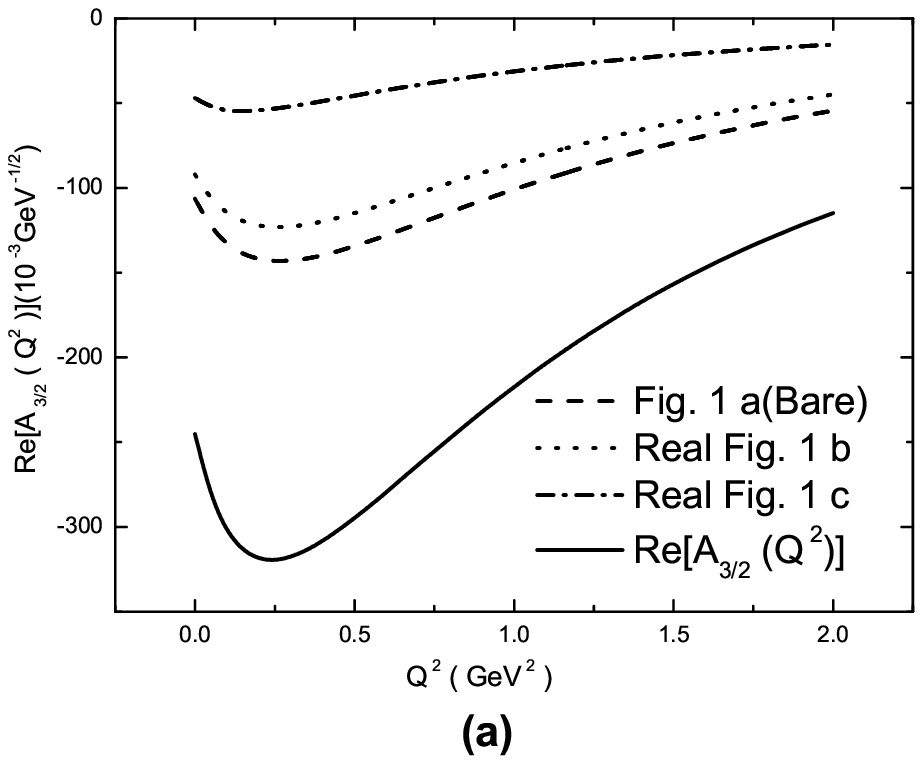}  %
\hspace{7mm}%
\includegraphics[width=75mm,height=75mm]{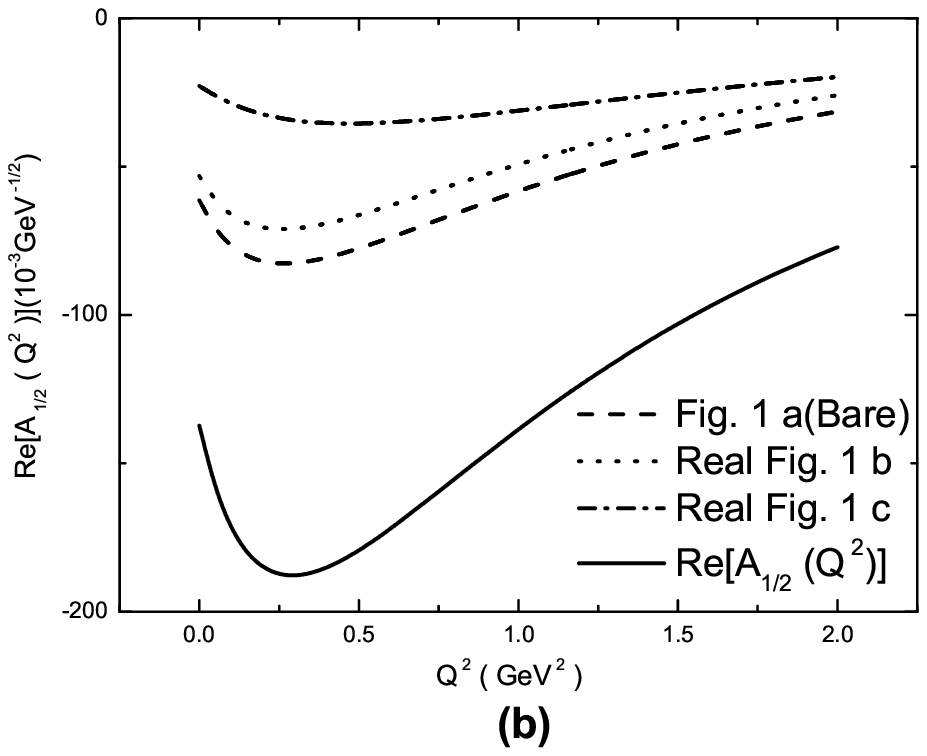}  %
\renewcommand{\figurename}{Fig.}
\caption{ The real part of the helicity amplitudes (a) $A_{3/2}$ and
(b) $A_{1/2}$ (b) of the $\Delta(1232)$ resonance with the different
contributions of Fig. \ref{feynman}. }
\label{helicity}  %
\end{figure}

%%%%%%%%%%%%%%%%%%%%%%%%%%%%%%%%%%%%%%%%%%%%%%%%%%%%%%%%%%%%%%%%%%%%%%%%%%
Figures \ref{helicity} (a) and \ref{helicity}(b) show the real part
of the individual contribution to the helicity amplitudes
$A_{3/2}(Q^2)$ and $A_{1/2}(Q^2)$ of the $\Delta(1232)$ resonance
respectively. As $Q^2=0$, the Hyc prediction of $A_{3/2}$
($A_{1/2}$) is $-243\times 10^{-3}~GeV^{-1/2}$ ($-135 \times
10^{-3}~GeV^{-1/2}$). Comparing it to the result $A_{3/2}=-187\times
10^{-3}~GeV^{-1/2}$ ($A_{1/2}=-108\times 10^{-3}~GeV^{-1/2}$)
without the pion meson cloud, we clearly see that the results with
the pion meson cloud effects reproduce the experimental data
$(-250 \pm 8) \times 10^{-3}~GeV^{-1/2}$ ($(-135 \pm 6) \times
10^{-3}~GeV^{-1/2}$) \cite{pdg} much better. It is clear that the
contributions from the pion cloud are significant in the real photon
point.

\par%

Another important observable in the $N-\Delta$ transition is the
ratio of $E2/M1$. Our result for it is -0.012. It reasonably agrees
with the experimental value reported in Particle data group
$-0.015\pm0.004$\cite{pdg}. If no meson cloud is considered, the
ratio of the simple hCQM (without the configuration mixing)
vanishes.  Our results show that the meson cloud effect also plays a
role on this ratio. Thus, we conclude that we  are able to
reproduce, at least partially, the experimental data of the
$N-\Delta$ transition in the low $Q^2$ region with hCQM and with the
pion meson cloud.

%%%%%%%%%%%%%%%%%%%%%%%%%%%%%%%%%%%%%%%%%%%%%%%%%%%%%%%%%%%%%%%%%%%%%%%%%%
\section{Summary}
\par\noindent\par%
In this work, we have studied the nucleon form factors and the form
factors of $\gamma^\star N\rightarrow\Delta$ transitions based on a
hCQM with Hyc wave function. The pion meson cloud effects are
explicitly included and discussed. From our numerical results, one
may conclude that the pion cloud contribution is crucial for a
reasonable explanation of the measured nucleon form factors as well
as of the helicity amplitudes of the $\Delta(1232)$ resonance.

\par%

Certainly, the pion meson cloud mainly affects the observables in the low
momentum transfer region. In the large momentum transfer region, we need
the hypercentral constituent quark model in the relativized version
\cite{hcqmff}, in addition to the inclusion of a perturbative
pion cloud. It will also be important to explore the effect of $SU(6)$
violating admixtures in the baryon wave functions \cite{isgur} simultaneously,
since the deformations of the nucleon and $\Delta$ wave functions can also
provide part of the E2/M1 ratio and part of the neutron charge form factor.

\par%

Finally, from the analysis of our results, we see that a
hypercentral constituent quark potential model ( with  a
hyper-coulomb plus a linear confinement term and the hyperfine term)
together with the pion cloud corrections might be able to give a
more reasonable description of the form factors of the nucleon and
of the transition form factors of the $\Delta(1232)$. It is of great
interest to see the pion meson cloud effects on the Roper resonance
and on the resonances $S_{11}(1535)$ and $D_{13}(1520)$ since the
transition amplitude to the $S_{11}(1535)$ resonance can be well
explained by the Hypercentral Constituent Quark Model in a large
$Q^2\sim 1~GeV^2$ region. This work is in progress.

\par%

\section*{Acknowledgments}
\par\noindent\par\vspace{0.2cm}
This work is supported by the Chinese National Science Foundations
(10475088, and 90103020), and by CAS Knowledge Innovation Project No.
KC2-SW-N02.

\par%

%%%%%%%%%%%%%%%%%%%%%%%%%%%%%%%%%%%%%%%%%%%%%%%%%%%%%%%%%%%%%%%%%%%%%%%%%

%%%%%%%%%%%%%%%%%%%%%%%%%%%%%%%%%%%%%%%%%%%%%%%%%%%%%%%%%%%%%%%%%%%%%%%%%

\end{document}